\documentclass[aps,prb,twocolumn,longbibliography,showpacs]{revtex4-1}
\usepackage{bm}
\usepackage{graphicx}
\usepackage{graphics}
\usepackage{amsmath}
\usepackage{amsfonts}
\usepackage{amssymb}
\usepackage{color}
\usepackage{xcolor}
\usepackage[unicode=true, bookmarks=false,breaklinks=false,pdfborder={0 0 1},backref=false,colorlinks=false]{hyperref}
\usepackage{makeidx}
\hypersetup{colorlinks=true,citecolor=blue,linkcolor=blue,filecolor=blue,urlcolor=blue}

\pdfoutput=1
\begin{document}

\newcommand{\be}   {\begin{equation}}
\newcommand{\ee}   {\end{equation}}
\newcommand{\ba}   {\begin{eqnarray}}
\newcommand{\ea}   {\end{eqnarray}}
\newcommand{\ve}  {\varepsilon}

\newcommand{\state} {\mbox{\scriptsize state}}
\newcommand{\band} {\mbox{\scriptsize band}}
\newcommand{\Dis} {\mbox{\scriptsize dis}}

\title{Impurity invisibility in graphene: \\ Symmetry guidelines for the design of efficient sensors}

\author{John Duffy}
\affiliation{School of Physics, Trinity College Dublin, Dublin 2, Ireland}
\author{James Lawlor}
\affiliation{School of Physics, Trinity College Dublin, Dublin 2, Ireland}
\author{Caio Lewenkopf}
\affiliation{Instituto de F\'{\i}sica,
Universidade Federal Fluminense, 24210-346 Niter\'oi RJ, Brazil}
\author{Mauro S. Ferreira}
\affiliation{School of Physics, Trinity College Dublin, Dublin 2, Ireland and \\ CRANN, Trinity College Dublin, Dublin 2, Ireland}

\date{\today}

\begin{abstract}
Renowned for its sensitivity to detect the presence of numerous substances, graphene is an 
excellent chemical sensor. 
Unfortunately, which general features a dopant must have in order to enter the list of substances 
detectable by graphene are not exactly known. 
Here we demonstrate with a simple model calculation implemented in three different ways 
that one of such features is the symmetry properties of the impurity binding to graphene. 
In particular, we show that electronic scattering is suppressed when dopants are bound 
symmetrically to both graphene sub-lattices, giving rise to impurity invisibility. In contrast, 
dopants that affect the two sublattices asymmetrically are more strongly scattered and therefore 
the most likely candidates to being chemically sensed by graphene. Furthermore, we demonstrate 
that impurity invisibility is lifted with the introduction of a symmetry-breaking perturbation such 
as uniaxial strain. In this case, graphene with sublattice-symmetric dopants will function as 
efficient strain sensors. We argue that by classifying dopants through their bonding symmetry 
leads to a more efficient way of identifying suitable components for graphene-based sensors.  
\end{abstract}


\maketitle

\section{Introduction}
\label{sec:introduction}

Due to its well-documented physical properties and numerous applications, graphene 
has been in the scientific limelight for over a decade now 
\cite{Geim2007,Geim2009,CastroNeto2009,yazyev_emergence_2010}. 
Due to its linear dispersion relation graphene shows some quite unique transport 
phenomena, such as Klein tunnelling, manifest as a suppression of backscattering
\cite{Mucciolo2010,Peres2010,DasSarma2011}. 
Here we focus on one of the more exciting features of graphene is the extreme sensitivity 
of its transport properties to relatively low disorder or impurity concentrations \cite{Li2008,Mucciolo2009,Robinson2008,Wehling2009,Yuan2010,Soriano2015}.
This makes graphene an attractive material for use in sensor-based applications, and indeed there 
has already been a lot of research in this direction, confirming its ability to detect substances 
at ultra-low concentrations (sub-PPM) \cite{Schedin2007,stine_real-time_2010,lu_gas_2009,robinson_reduced_2008,shao2010,He2012,Wu2013,Liu2015}. 

Which substances can graphene detect is a question that continues driving the search 
for atoms and molecules that impact its transport properties. This search has been mainly 
based on trial and error, {\it i.e.}, by exposing graphene to a variety of dopants in the hope
that they function as strong scattering centres \cite{Leenaerts2008,gorjizadeh_spin_2008,Wehling2009,Wehling2010,Katoch2015,Herrero2016}. 
Owing to the overwhelming number of possibilities to account for, it is no surprise that this 
{\it ad-hoc} approach fails to provide insight on the conditions that ideal dopants must have to
make good graphene-based sensors. Rather than trial and error, a more general approach 
is needed to guide the search for efficient sensors. 

With that goal in mind, this study makes use of a simple model calculation that describes the 
electronic scattering in impurity-doped graphene. Rather than specifying the exact form 
and detailed characteristics of the doping impurities, we adopt a more general approach 
that aims to separate the distinct contributions to scattering events: one that depends on 
the intrinsic specificity of dopants and another that is determined primarily by their bonding 
symmetry. Whereas there is an enormous variety of impurities that interact with graphene, 
there are only a few different conformations that characterize the bonding symmetry. Remarkably, 
out of this small number of symmetries, we show that there is one in particular that gives rise 
to vanishingly small scattering regardless of the specific details of the dopant. 
This class of dopants is therefore expected to be  hardly visible for the conduction electrons. 

This finding corresponds to a considerable advance from the aforementioned {\it ad-hoc} strategy since 
we are able to infer about the graphene properties of a whole range of dopants that have this 
particular bonding symmetry. Most importantly, because the predicted weak-scattering behaviour is 
symmetry-dependent, any symmetry-breaking perturbation is likely to enhance the scattering 
strength of this class of impurities. Therefore, we argue that graphene doped with impurities 
that have this particular bonding symmetry will give rise to devices that are extremely sensitive 
to, for instance, uniaxial mechanical strain. 

Regarding the sequence adopted in this manuscript, we start by defining the model 
Hamiltonian, followed by a few different yet complementary ways of accounting for 
the scattering contribution of impurities in graphene. All these approaches lead to the 
same conclusion, {\it i.e.}, that impurities with certain bonding symmetries may be 
completely transparent, causing hardly any electronic scattering. We finish by 
discussing possible consequences that this feature might bring to the field of 
sensor design. 

\section{Model Hamiltonian}
\label{sec:model}

Let us define the model Hamiltonian used throughout the manuscript. The system consists of 
a graphene sheet with one single impurity described by the Hamiltonian $H = H_0 + V$, 
where 
\begin{equation}
H_0 = -\sum_{\langle i,j \rangle} \vert i \rangle t_{i,j} \langle j \vert + 
\sum_{\langle \ell, m \rangle} \vert \ell \rangle h_{\ell,m} \langle m \vert 
\end{equation}
corresponds to the pristine nearest-neighbour tight-binding graphene Hamiltonian 
defined by the matrix elements $t_{i,j}$ plus a single impurity defined by the matrix 
elements $h_{\ell,m}$. The indices $i$ and $j$ label the graphene sites while 
$\ell$ and $m$ label the impurity sites. The states $\vert \alpha \rangle$ represent an 
atomic orbital centred at site $\alpha$, where $\alpha = \{i,j,\ell,m\}$. The matrix 
elements $t_{i,j} = t$ when $i$ and $j$ are nearest neghbours and vanish otherwise. 
The value of $t=2.7$  eV is known to reproduce well the low energy electronic structure 
of graphene and will hereafter serve as our energy unit.  Although not specified, 
the matrix elements $h_{\ell,m}$ may describe a variety of possible impurities ranging 
from single atoms to more complex structures such as molecules and nanoparticles. 
The only assumptions made about the impurity is that it connects to graphene through 
only one of its sites (labelled $\ell=0$) and that the orbital $\vert 0 \rangle$ possesses 
certain symmetries. These assumptions can be easily relaxed, should we need to 
consider impurities of a more complex geometrical structure and/or with orbitals of 
different symmetries \cite{Uchoa2011,Garcia2014}.   

\begin{figure}[h]
\begin{center}
\includegraphics[width=0.55\columnwidth]{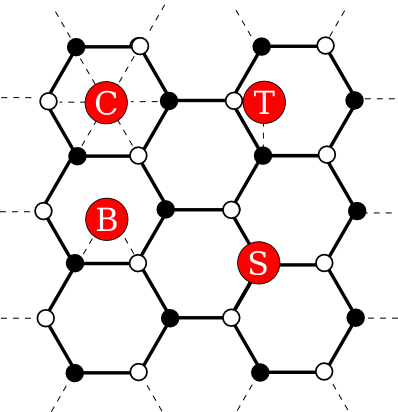}\\
\caption{(Color online) Schematic representation of a centre-bonded (labelled with a C); 
a top-bonded (labelled with a T); and a bridge-bonded (labelled with a B) impurity. 
For the sake of completeness, we also include a schematic representation of a 
substitutional impurity (labelled with an S). Impurities are represented by red circles while 
filled and hollow circles correspond to the two graphene sublattices also referred to 
as $A$ and $B$ sublattices.}
\label{fig:impurity_position}
\end{center}
\end{figure}

Note that $H_0$ describes a graphene sheet and a single impurity totally decoupled from one 
another. The graphene-impurity coupling term is described by $V$ and depends on the bonding 
conformation. We assume that impurities can be either centre-,  top- or bridge-bonded to the 
graphene lattice, as schematically depicted in Fig.~\ref{fig:impurity_position}. For the sake 
of completeness, we also include substitutional impurities in the figure because their scattering 
response is practically identical to top-bonded impurities.

The coupling operator $V$, now renamed $V_T$, $V_B$ and $V_C$ depending on 
the conformation type, is defined as
\be
V_T =   |0\rangle \tau \langle 1|  + {\rm H.c.}
\ee
\be
V_B =   |0\rangle \tau_1 \langle 1|  +  |0\rangle \tau_2 \langle 2| + {\rm H.c.}
\ee
\be
V_C =  \sum_{i \in {\rm R}}   |0\rangle \tau_i \langle i| + {\rm H.c.}\,\,,
\ee
where the subscripts $T$, $B$ and $C$ refer to top, bridge and centre, respectively. 
The state $\vert 0 \rangle$ represents an orbital centred at the impurity site that is bonded 
to the graphene sheet, whereas the others are orbitals centred on graphene sites.  
In the center bonded case, the sum runs over the six carbon sites of 
the hexagonal ring R surrounding the impurity.

In the case of centre-bonded impurities, the values of $\tau_i$ depend on how the impurity 
hybridizes with the graphene atoms. The possibilities are \cite{Uchoa2011,Garcia2014}: 
(i) It may hybridize equally with all six neighboring carbon atoms ($\tau_i = \tau$), which 
is the case for $s$ and $d_{z^2}$ orbitals; (ii) It may have a $\pi$-phase difference in the 
hybridization of the adsorbed impurity with the two different sublattices, 
($\tau_1 = \tau_3=\tau_5 =-\tau_2=-\tau_4=-\tau_6\equiv \tau$), which is typical of $f$ orbitals; 
(iii)  For a $d_{xy}$ orbital, $\tau_1 = \tau_4=0$ and $\tau_2=\tau_5 =-\tau_3=-\tau_6=
\sqrt{3} \tau/2$; 
(iv) For a $d_{x^2-y^2}$ orbital, $\tau_1 = \tau_4=\tau$ and
$\tau_2=\tau_3=\tau_5=\tau_6=-\tau/2$. Here we focus on the first two possibilities, where 
the values of $\vert \tau_i \vert$ are the same for all $i$. 

\section{Impurity scattering}
\label{sec:scattering}

Having defined the model Hamiltonian, three distinct approaches will be used to study how the 
graphene conductance is impacted by the introduction of different bonding-symmetry impurities. 
First we investigate how the scattering events are described by the real-space Green functions 
of impurity-doped graphene. We then turn our attention to writing the scattering cross section 
in wave-number domain and analyze how it is affected by the introduction of graphene dopants. 
Finally, the conductance of doped graphene is numerically calculated through the Kubo formula. 
These complementary methods shed light on different aspects of the scattering process and how the 
electronic transport is affected by the bonding-symmetry of the dopants. 

\subsection {Scattering in real space}
\label{sec:real_space}

It is convenient to describe the scattering processes associated to adsorbed impurities in graphene in 
terms of the $T$-matrix. The latter is defined by $G = g + g T g$, where $G = (E - H)^{-1}$ 
the system full Green's function and $g=(E - H_0)^{-1}$ is the free Green's function. The $T$-matrix 
can be obtained from the Dyson equation, $G = g + g \, V \, G$, using standard Green's function 
techniques \cite{Economou06}.

This strategy allows us calculate the electronic propagator between two arbitrary graphene sites 
$a$ and $b$ in the presence of the graphene-impurity coupling term. The simplest case to consider 
is of top-bonded impurities described by $V_T$. Here,
\be
G_{a,b} = g_{a,b} + g_{a,1 } \, T_{T} \, g_{1,b}\,\,,
\label{Gab_top}
\ee
where 
\be
T_{T} = \Sigma (1 - g_{11}\Sigma)^{-1}
\ee
is the relevant $T$-matrix element and 
\be
\Sigma \equiv \vert \tau \vert^2 \, g_{0,0},
\label{self}
\ee 
where $g_{0,0}$ is the uncoupled impurity Green's function projected on $|0\rangle$. 
$\Sigma$ acts as the self-energy associated with the impurity. 

Scattering is fully described by the second  term on the r.h.s.~of Eq.~\eqref{Gab_top}.
It is worth noting that the part of the $T$-matrix  that depends on the specific details of 
the impurity is entirely contained in the self-energy, {\it i.e.}, in the ``contact" Green function 
$g_{0,0}$. Therefore, scattering caused by any top-bonded impurity is fully taken into 
account by Eq.~\eqref{Gab_top} once the Green function $g_{0,0}$ is known.  

Similar steps are followed to obtain the propagator $G_{a,b}$ for bridge- and 
centre-bonded impurities, taking care to replace $V_T$ by $V_B$ and $V_C$, 
respectively. The corresponding full Green's functions are given by 
\be
G_{a,b} = g_{a,b} + (g_{a,1 }+g_{a,2}) \, T_B \, (g_{1,b}+g_{2,b})\,\,,
\label{Gab_bridge}
\ee
and 
\be
G_{a,b} = g_{a,b} + (g_{a,1 }+...+g_{a,6}) \, T_C \, (g_{1,b}+...+g_{6,b})\,\,,
\label{Gab_centre}
\ee
respectively. 
The $T$-matrix for the bridge- and the centre-bonded impurities are denoted, respectively, 
by  $T_B$ and $T_C$. They read
\be
T_B =  \Sigma \, (1 - \gamma_B \, \Sigma)^{-1} \, 
\ee
and 
\be
T_C =  \Sigma \, (1 - \gamma_C \, \Sigma)^{-1} \,\,,
\label{Tc}
\ee
where $\gamma_B = \sum_{i,j=1}^2 g_{ij}$ and $\gamma_C = \sum_{i,j=1}^6 g_{ij}$ are 
sums of all the matrix elements of $g$ involving graphene sites that are bonded to the 
impurity. $\gamma_B$ involves a sum over 4 terms and $\gamma_C$ a sum over 36 terms. 

A simple analysis of Eq.~\eqref{Gab_top} indicates that, for top-bonded impurities, the
scattering contribution to the propagator $\Delta G_{a,b} \equiv G_{a,b}-g_{a,b}$ contains 
the usual product of three quantities: (i) one off-diagonal propagator $g_{a,1}$ between 
site $a$ and the scattering site on graphene, (ii) the relevant matrix element of the 
$T$-matrix and (iii) another off-diagonal propagator $g_{1,b}$ this time associated 
with site $b$. In this case, the only way the scattering can be weak is if the $T_T$ itself 
is small, {\it i.e.}, if the top-bonded impurity is a weak scatterer. 

The situation changes for bridge- and centre-bonded impurities, described respectively by 
Eqs.~(\ref{Gab_bridge}) and (\ref{Gab_centre}). The scattering correction of the propagator 
is still written as a product of three separate terms, one of which being the $T$-matrix. The other two 
terms involve not one single propagator but a sum of several propagators $g$. More specifically, 
the sums $\alpha_B = g_{a,1} + g_{a,2}$ and $\beta_B = g_{1,b}+g_{2,b}$ appear in 
Eq.~\eqref{Gab_bridge} while the sums $\alpha_C = g_{a,1} + ... + g_{a,6}$ and 
$\beta_C = g_{1,b}+ ... +g_{6,b}$ can be seen in Eq.~\eqref{Gab_centre}. 

\begin{figure}[ht]
\begin{center}
\includegraphics[width=1.0\columnwidth]{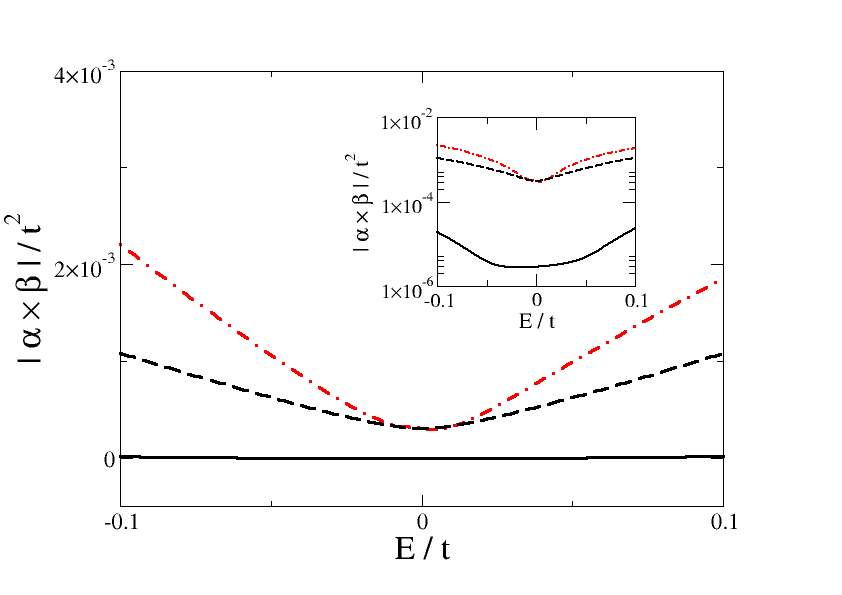}\\
\caption{(Color online) Scattering contribution to the single-particle Green function 
$\vert \alpha \times \beta \vert$ (in units of $t^{-2}$) as a function of the energy (in units of 
the electron hopping $t$). The (black) solid line corresponds to the 
case of centre-bonded impurities described by Eq.~\eqref{Gab_centre}; 
the (black) dashed line is for bridge-bonded impurities described by 
Eq.~\eqref{Gab_bridge} and the (red) dot-dashed line represents the 
case of top-bonded impurities according. Note that in the case of top-bonded 
mpurities the product $\alpha \times \beta = g_{a,1} \times g_{1,b}$. The inset 
shows the same results in a linear-log plot.}
\label{fig:interference}
\end{center}
\end{figure}

The sums in $\alpha_C$ and $\beta_C$ give rise to interference effects that
strongly modify $\Delta G_{ab}$.
To investigate the magnitude of the interference contribution, we show in 
Fig. \ref{fig:interference} results for the product $\vert \alpha_B \times \beta_B \vert$ 
and $\vert \alpha_C \times \beta_C \vert$, for bridge- and 
centre-bonded impurities, respectively. 
These results are for impurities halfway between sites $a$ and $b$, chosen to be 
a distance $20 \, a_0$ apart along the armchair direction, where $a_0$ is the graphene 
lattice parameter. Sites $a$ and $b$ are arbitrarily chosen carbon atoms on the 
graphene lattice and the results shown in Fig.~\ref{fig:interference} re not qualitatively 
affected by any specific choice of their values. For the sake of comparison, we also 
plot the equivalent product $\vert g_{a,1} \times g_{1,b} \vert$ seen in Eq.~(\ref{Gab_top}) 
for the case of top-bonded impurities. 
Results are plotted as a function of the electron energy. While all curves have a minimum at the 
Dirac point, the most revealing aspect of this figure is what happens to the curves as the energy 
moves away from $E=0$. Results for top- and bridge-bonded impurities increase fairly rapidly, 
but the centre-bonded case is remarkably different.  The quantity $\alpha_C \times \beta_C$ is 
orders of magnitude smaller than the other two cases and clearly indicates that the sums 
$(g_{a,1 }+...+g_{a,6})$ and $(g_{1,b}+...+g_{6,b})$ vanish as a result of destructive interference
 in the propagators $g$. The log-scale plot in the inset of 
Fig.~\ref{fig:interference} shows that the centre-bonded results are between 2 and 3 orders 
of magnitude weaker than the other two cases. The same could have been concluded by 
using the analytical expression for the off-diagonal matrix elements of the graphene Green's 
function within the tight-binding approximation
\cite{Power2011}.

 A direct consequence of the results shown in Fig.~\ref{fig:interference} is that  centre-bonded 
 impurities, except for the $d$-orbital ones, are very weak scatterers. The generality of our 
 argument is based on the 
 fact that even without specifying what impurities are being considered, the destructive interference 
 experienced by the propagators in Eq.~(\ref{Gab_centre}) will give rise to results that are orders 
 of magnitude smaller than those obtained by Eqs.~(\ref{Gab_top}) and Eq.~(\ref{Gab_bridge}). 
 Note that this effect occurs regardless of the value of the real-space $T$-matrix $T_C$, unless of 
 course the T-matrix has resonance levels at very specific energies. We address the issue of 
 resonances in Sec.~\ref{sec:Kubo}. 

\subsection{Scattering cross section}
\label{sec:cross_section}
 
We now turn our attention to the cross section of graphene in the presence of top-bonded 
($\sigma_T$) and centre-bonded ($\sigma_C$) impurities. Results for bridge-bonded impurities 
will not be pursued simply because, as seen in Fig.~\ref{fig:interference}, they behave very 
similarly to the case of top-bonded dopants. 

The cross section is directly related to the $T$-matrix. 
In scattering theory, the $T$-matrix is usually defined 
\be
V |\psi_E^{(+)}\rangle = T^+(E) | {\bf k}\rangle,
\ee
where $|\psi_E^{(+)}\rangle$ is a solution of the Lippmann-Schwinger equation,
namely
\be
|\psi_E^{(+)}\rangle =  |{\bf k} \rangle + g^+\!(E) V  |\psi_E^{(+)}\rangle.
\ee
We are interested in the scattering amplitude $f_{\rm scat}({\bf k}, {\bf k} ') \sim
 \langle {\bf k} | T | {\bf k} '\rangle$, that is directly related to the scattering cross-section 
 and the transport time that appears in the Boltzmann equation used in the analysis of
 the transport properties of graphene in the diffusive regime 
 \cite{CastroNeto2009,Mucciolo2010,Peres2010}. 
The fundamental difference to the previous section is that here 
we express all key quantities in the wave-number basis, as opposed to the real-space basis. 

Recalling that the tight-binding Hamiltonian for pristine graphene can be written in 
momentum representation as
\be
H_{0,{\rm graphene}} = t \left(
\begin{array}{cc}
0 & f({\bf k}) \\
f^*({\bf k}) & 0 \end{array}
\right),
\ee
where 
\be
f({\bf k}) =  - (e^{-i {\bf k}\cdot {\bf a}_1} + 
e^{-i {\bf k}\cdot {\bf a}_2} + 1).
\ee
The corresponding eigenstates read \cite{Bena2009}
\be
| {\bf k,\pm} \rangle = 
\frac{1}{\sqrt{2}}\sum_n e^{i {\bf k}\cdot {\bf R}_n}  
\left(| n, A\rangle \pm 
e^{i \theta({\bf k})} | n, B\rangle \right),
\ee
where $+$ ($-$) corresponds to positive (negative) eigenenergies, ${\bf R}_n^{}={\bf R}_n^A$, 
and $\theta({\bf k}) = {\rm arg} [f({\bf k})]$. $A$ and $B$ refer to the two equivalent graphene 
sublattices.  Note that $| {\bf k,\pm} \rangle$ are scattering states normalized to a Dirac delta-function,
and, consequently, that have a different normalization than the states defined in 
Ref.~\onlinecite{Bena2009}.

\begin{figure}[h!]
\begin{center}
\includegraphics[width=0.65\columnwidth]{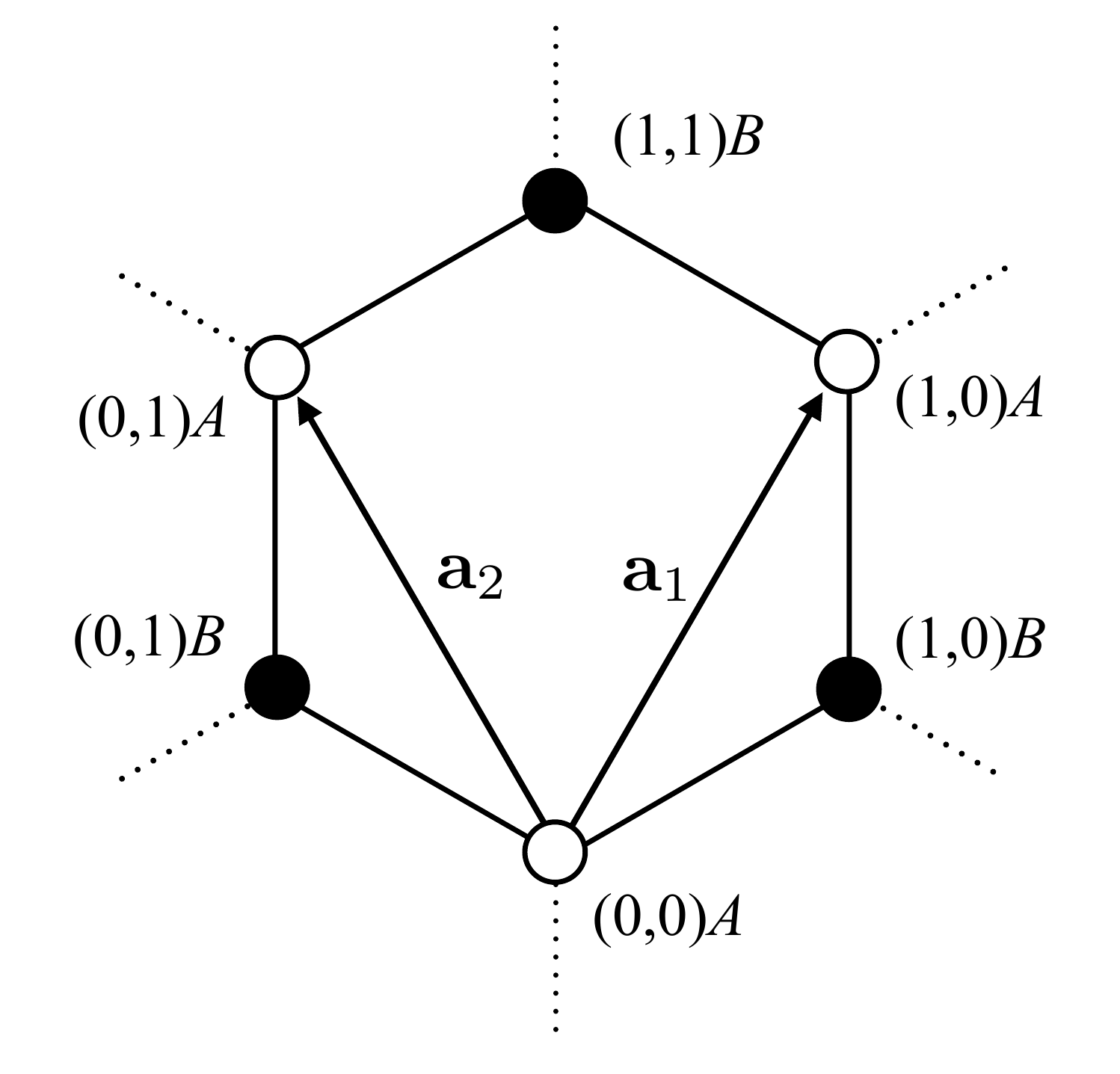}
\caption{Schematic representation of the lattice labels $i=(m,n)$ corresponding
to ${\bf R}_i = m {\bf a}_1 + n{\bf a}_2$. For the top-bonded case, the impurity is
placed atop of the (0,0) site, whereas for the center-bonded case, the impurity is 
at the center of the hexagonal ring. }
\label{fig:lattice_labels}
\end{center}
\end{figure}

Since $|{\bf k}\rangle$ distinguishes between $A$ and $B$ sites, we introduce the lattice 
labelling defined in Fig.~\ref{fig:lattice_labels}. The primitive unit cell consists of a pair of 
$A$ and $B$ sites connected by a vertical bond. The PUCs are defined as $i=(m,n)$ 
with ${\bf R}_i = m {\bf a}_1 + n{\bf a}_2$.

For top-bonded impurities, the explicit representation of the $T$-matrix operator in 
the site basis is
\be
\label{eq:Tcenter}
T_T = |i\rangle \langle i| T_T |i\rangle \langle i| = {\cal T}_T(E) \sum_{i} |i\rangle  \langle i |.
\ee
For a single top-absorbed impurity, we write
\begin{align}
\langle {\bf k}', &\pm | T_T | {\bf k},\pm \rangle = 
\frac{1}{2} \sum_{i,j}  e^{-  i{\bf k}'\cdot {\bf R}_i+ i{\bf k}\cdot {\bf R}_j}
\nonumber\\&\times
\left( \langle i,A| \pm e^{i \theta({\bf k}')} \langle i,B| \right) 
T_T
\left( |j,A\rangle  \pm e^{-i \theta({\bf k})} | j,B\rangle \right) .
\end{align}
Hence, if $i$ belongs to the $A$ sublattice we obtain
\begin{align}
\langle {\bf k}', \pm | T_T | & {\bf k},\pm \rangle = 
\frac{ {\cal T}_T(E)}{2}  e^{-  i({\bf k}'-{\bf k})\cdot {\bf R}_i}
\end{align}
while if $i$ belongs to the $B$ sublattice 
\begin{align}
\langle {\bf k}', \pm | T_T | & {\bf k},\pm \rangle = 
\frac{  {\cal T}_T(E) }{2} e^{-  i({\bf k}'-{\bf k})\cdot {\bf R}_i} e^{i \theta({\bf k}')-i \theta({\bf k})}.
\end{align}
 
The results for $T_T$ are then further simplified by identifying the $i$ site as $(0,0)$ in 
the notation of Fig.~\ref{fig:lattice_labels}. Note that if one is interested in coherent multiple 
scattering due to a finite concentration of impurities, the 
relative phases are important. Let us restrict ourselves to the low impurity concentration, where 
coherent multiple scattering is unlikely to play a significant role.

As a result,
\begin{equation}
\langle {\bf k}', \pm | T_T |  {\bf k},\pm \rangle = \frac{ {\cal T}_T(E)}{2} 
\end{equation}
for $i\in A$ and 
\begin{equation}
\langle {\bf k}', \pm | T_T |  {\bf k},\pm \rangle =  \frac{ {\cal T}_T(E)}{2}  e^{i \theta({\bf k}')-i \theta({\bf k})}
\end{equation}
for $i\in B$. The difference is just a phase, which is immaterial for the cross section. 
Despite not affecting the cross section for top-bonded dopants, this phase difference 
will have a dramatic effect when impurities are coupled equally to both $A$- and 
$B$-sublattices, as we demonstrate next. 


Following similar steps, we now derive the wave-vector dependent $T$-matrix associated with 
centre-bonded impurities. For a single-impurity this quantity is expressed in the real-space basis by 
\begin{align}
\label{eq:Tcenter2}
T_C =  \sum_{(i,j)\in R} |i\rangle \langle i| T_C |j\rangle \langle j| 
 = 
{\cal T}_C(E) \sum_{(i,j)\in R} |i\rangle  \langle j |.
\end{align}

Hence, for an impurity centred at the $R$th hexagon, one has
\begin{align}
\langle {\bf k}',\pm | & T_C | {\bf k},\pm\rangle = 
\frac{1}{2} \sum_{i,j \in R}  e^{-  i{\bf k}'\cdot {\bf R}_i+ i{\bf k}\cdot {\bf R}_j}
\nonumber\\&\times
\left( \langle iA| \pm e^{i \theta({\bf k}')} \langle iB| \right) \!
T_C \!
\left( |jA\rangle  \pm e^{-i \theta({\bf k})} | jB\rangle \right) .
\end{align}

\begin{widetext}

The above expression yields 36 terms to compute. After a long, but straightforward calculation,
one obtains
\begin{align}
\langle {\bf k}',+ | T_C | {\bf k},+\rangle = 
 & \frac{{\cal T}_C(E) }{2}
 \left[  f({\bf k}') + e^{i\theta({\bf k}')}  e^{-i {\bf k}\cdot( {\bf a}_1+ {\bf a}_2)}  f^*({\bf k}') \right]
 \left[  f^*({\bf k}) + e^{-i\theta({\bf k})}  e^{i {\bf k}'\cdot( {\bf a}_1+ {\bf a}_2)}  f({\bf k}) \right],
\end{align}
for positive energies.
\end{widetext}

For low energies, it is convenient to expand the wave vectors around the $K$-points
\cite{CastroNeto2009}, namely,  ${\bf k} = {\bf K}^\xi + {\bf q}$ and  ${\bf k}' = 
{\bf K}^\xi + {\bf q}'$, where $\xi = \pm$ is the valley index.
For what follows it is useful to recall that 
\be
{\bf K}^\xi \cdot {\bf a}_1 = \frac{2\pi}{3} \xi 
\quad \mbox{and} \quad
{\bf K}^\xi \cdot {\bf a}_2 = -\frac{2\pi}{3} \xi \;.
\ee
Note that ${\bf k}\cdot({\bf a}_1+ {\bf a}_2) = {\bf K}^\xi \cdot({\bf a}_1+ {\bf a}_2) + 
{\bf q} \cdot({\bf a}_1+ {\bf a}_2) =  3aq_y$.

Expanding $f({\bf k})$ to first order in $\bf q$, one obtains
\be
f_\xi({\bf k}) = \frac{3}{2} qa\, e^{-i \theta_\xi({\bf q})}
\ee
where
\be
e^{-i \theta({\bf k})} \equiv e^{-i \theta_\xi({\bf q})} = \xi \frac{q_x}{q} + i \frac{q_y}{q}.
\ee

We are now ready to conclude the calculation of the $k$-dependent $T$-matrix, namely
\begin{align}
\langle {\bf q}' \xi' +& | T_C | {\bf q}\xi+\rangle =   {\cal T}_C(E) \frac{9a^2}{4} qq' 
e^{ i[\theta_{\xi'}({\bf q}') - \theta_{\xi}({\bf q})]/2}
\nonumber\\
 \times &  
\Big\{ \!\cos 3[\theta_{\xi'}({\bf q}') - \theta_{\xi}({\bf q})] +  \cos 3[\theta_{\xi'}({\bf q}') +\theta_{\xi}({\bf q})]
\Big\},
\label{Tk}
\end{align}
where $|{\bf k}+\rangle \equiv | {\bf q}\xi+\rangle$.

Two immediate conclusions can be extracted from Eq.~(\ref{Tk}): 
(i) The angular part of the scattering cross section displays the familiar $2\pi/3$-periodicity that 
is inherent to the hexagonal symmetry of graphene; 
(ii) the $T$-matrix scales as $(qa)^2$ for low energies for both intra ($\xi=\xi'$) and intervalley 
($\xi \neq \xi'$) scattering. The latter conclusion reiterates the results shown in 
Sec.~\ref{sec:real_space} and demonstrates once again that centre-bonded 
impurities hardly affect the transport properties of electrons near the Dirac point.  

It is worth emphasizing that this is a situation where a short-range impurity, taking into  account
intra- and intervalley scattering process, is suppressed by interference effects. This is quite different from 
the standard picture inferred from the scattering analysis of the Dirac equation in graphene, where one 
identifies the suppression of backscattering with long-range impurities and as a manifestation of Klein 
tunneling \cite{Mucciolo2010}.

\subsection{Numerical results}
\label{sec:Kubo}

We now study resonance scattering regime and the case of finite impurity-doped graphene 
systems, that involve multiple scattering. 
For that purpose we numerically calculate the conductance using the Kubo formula. The zero bias conductance $\Gamma$ reads  \cite{lee_anderson_1981,mathon_oscillations_1997,Costa2013}
\be
\!\!\!
\Gamma = \frac{4 e^2}{\hbar} \rm{Re} \rm{Tr} \left[ \tilde{\bf G}_{00} {\bf U}_{01} \tilde {\bf G}_{11} {\bf U}_{10} 
- {\bf U}_{01} \tilde {\bf G}_{10} {\bf U}_{10} \tilde {\bf G}_{10} \right]
\ee
where $\tilde {\bf G}_{j,\ell} = ( {\bf G}_{j,\ell}^- - {\bf G}_{j,\ell}^+ )/2i$ is the difference 
between the retarded and advanced Green functions. Here, ${\bf G}^\pm_{j,\ell}$ is the
matrix formed by the Green function elements connecting
unit cells $j$ and $\ell$. For computational purposes it is convenient  define a 
finite-sized graphene sample across which the conductance is calculated.  
In practice, this comes down to defining three graphene nanoribbons of similar width, 
two of which are semi-infinite and act as leads, separated by a finite-length section where 
impurities are placed. 
The indices 0 and 1 correspond the the interface between the central region and one of 
the leads.
By taking 
the boundary conditions properly, this method gives the same results as the standard 
recursive Green's function method \cite{Lewenkopf2013}. Unit cells are defined as lines 
across the ribbon width and the trace is taken over both site and spin indices. 
Similarly, ${\bf U}_{j,\ell}$ represents a matrix consisting of off-diagonal hopping terms 
connecting neighboring unit cells $j$ and $\ell$.  
All Green functions above are evaluated at $E_F$.

Rather than treating the scatterers as general objects described by the self-energy 
definition of Eq.~(\ref{self}) as before, we must now specify the impurity in order to 
evaluate the conductance. We consider impurities with on-site energy $\epsilon_a$, 
which modifies the self-energy $\Sigma$ to
\be
\Sigma = \vert \tau \vert^2 \, (E - \epsilon_a)^{-1}\,.
\ee 

\begin{figure}[ht]
\begin{center}
\includegraphics[width=0.95\columnwidth]{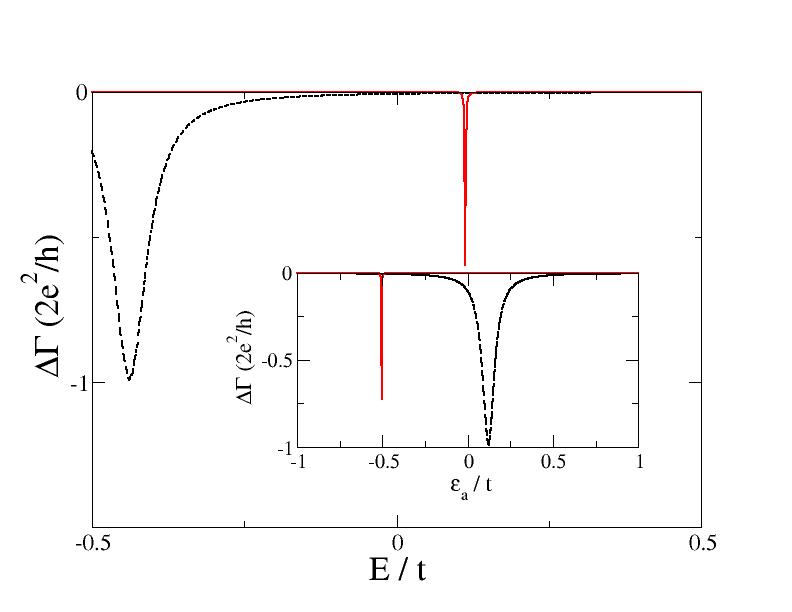}\\
\caption{(Color online) Change in the conductance $\Delta \Gamma$ (in units of $2 e^2/h$) 
as a function of the Fermi energy for a single impurity with on-site energy $\epsilon_a = -0.5 \, t$,
and $\tau = -0.5 \, t$. 
Solid red (dashed black) line corresponds to the case of a centre-bonded (top-bonded) impurity. 
Inset depicts the same quantity for a fixed Fermi energy $E_F=0.11 \, t$, this time plotted as a 
function of the impurity's on-site energy.}
\label{fig:kubo}
\end{center}
\end{figure}

Figure \ref{fig:kubo} shows numerically evaluated results for the change in the conductance 
of a 8-atom wide graphene ribbon due to the presence of a single impurity as a function of 
the Fermi energy. Solid (dashed) line is for a centre-bonded (top-bonded) 
impurity. Top-bonded impurities are strong scatterers when compared to their centre-bonded 
counterparts. The dashed line shows that the conductance with top-bonded impurities is 
significantly reduced across a wide range of energies. 

In contrast, the conductance of graphene with centre-bonded impurities is practically identical 
to that of the pristine case, except for a very narrow energy range around the resonance. This 
corroborates the preceding argument that the centre-bonded symmetry makes impurities with 
this type of bonding  hardly visible to the conduction electrons. The narrow peak seen in 
the main panel of Fig.~\ref{fig:kubo} is explained by Eq.~\eqref{Gab_centre}. While the quantity 
$\alpha_C \times \beta_C$ approaches zero, as seen in Fig. \ref{fig:interference}, it is possible 
to find a suitable Fermi energy that leads to $\gamma_C \times \Sigma \approx 1$. When that 
happens the $T$-matrix $T_C$ in Eq.~(\ref{Tc}) diverges, compensating the destructive interference
effects. However, this is practically accidental and calls for some fine tuning of the 
Fermi energy and/or of the impurity resonance values, neither of which are very practical. 

To make this point more explicitly, in the inset of Fig.~\ref{fig:kubo} we have also show $\Delta \Gamma$
plotted as a function of the impurity on-site energy for a fixed $E_F$.
Once again, an extremely narrow isolated peak suggests that the lack of transparency of 
centre-bonded impurities is not a robust feature but results from a coincidental match of energies. 

In fact, in a recent paper, Garc\'{\i}a and collaborators \cite{Garcia2014} studied the Anderson 
localization driven by adatom disorder in graphene. They find that despite the suppression of the 
scattering cross section due to  destructive interference in center adsorbed impurities, the system 
undergoes an Anderson metal-insulator transition but only for particular values of the doping and 
the impurity resonance energy. Our results evidence that such a transition requires a very precise 
parameter tuning.

\begin{figure}[h]
\begin{center}
\includegraphics[width=1.\columnwidth]{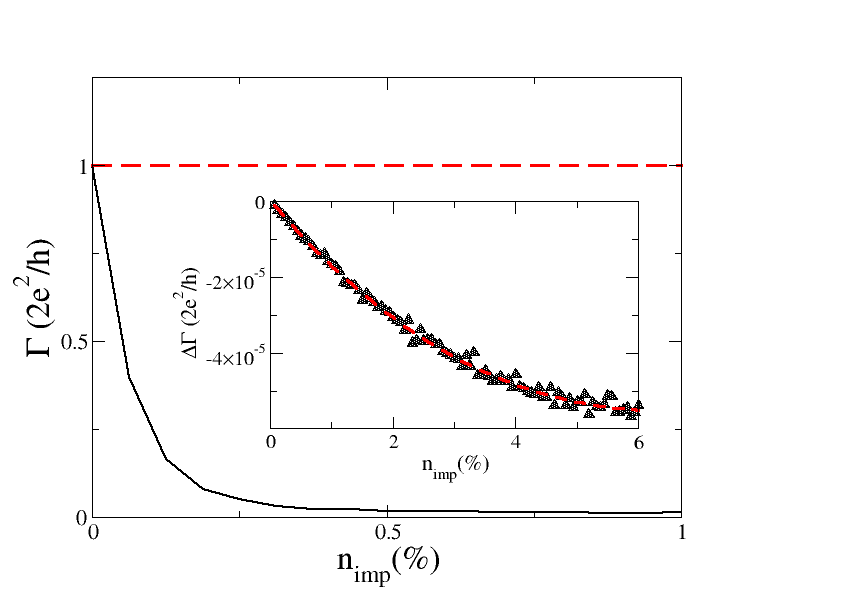}\\
\caption{Conductance (in units of $2 e^2/h$) as a function of the impurity 
concentration $n_{\rm imp}$. Solid (dashed) line corresponds to the case of top-bonded 
(centre-bonded) impurities. $E_F=0$, $\tau = -2 \, t$ and $\epsilon_a = 0.3 \, t$. Inset shows the change in conductance  
$\Delta \Gamma$ , indicating that the conductance actually decreases with increasing $n_{\rm imp}$, but extremely slowly. All points were averaged over 1000 configurations. }
\label{fig:kubo2}
\end{center}
\end{figure}

Instead of considering single impurities, we now proceed to studying how the conductance 
$\Gamma$ changes as the impurity concentration $n_{\rm imp}$ increases, as shown in 
Fig.~\ref{fig:kubo2}. Extensive configurational averaging was carried out to obtain statistical 
significance in our calculations. A very small percentage (0.01\%) of top-bonded impurities 
(dashed line) is sufficient to reduce the conductance of a graphene ribbon to 50\% of its 
pristine value whereas no reduction can be seen for centre-bonded scatterers (solid line). 
To perceive any reduction in the case of centre-bonded impurities, concentrations must 
exceed the 5\% mark and yet the reduction is orders of magnitude smaller than that seen 
for the top-bonded symmetry case. The inset shows the change in conductance for 
centre-bonded impurities and reductions are practically negligible, confirming once again 
our ideas of impurity invisibility. When searching for experimental signatures of the distinct 
responses of top- and centre-bonded impurities, the results of Fig.~\ref{fig:kubo2} are the most evident.

\section{Discussion and Conclusions}

Having demonstrated that the centre-bonded symmetry indeed gives rise to impurity transparency, it is worth now discussing what repercussions this finding brings. For a start, one may conclude that graphene can function as a good sensor of substances whose contact to the underlying hexagonal structure resembles that of top-bonded impurities. More specifically, it should be sensitive to the presence of impurities  whose binding to the graphene sublattices is asymmetric, {\it i.e.}, the effect that the impurity causes to the graphene sublattices is different. As shown in Figs.~\ref{fig:interference}, \ref{fig:kubo}, and \ref{fig:kubo2}, the scattering caused by top-bonded impurities is always the largest of all analyzed cases (for the same hybridization matrix element). This in itself is a valuable finding since it may offer a clear guideline in the search for substances that graphene can detect instead of the common {\it ad-hoc} approach of trial and error. 

Another consequence is that centre-bonded impurities are  not ideal 
for generating chemical sensors since they may be 
transparent. Nevertheless, they may be employed 
in the construction of sensors of a different nature. Because the impurity transparency results 
from a symmetry-driven destructive interference in the scattering cross section, all one needs 
to do to turn transparent objects into opaque scatterers is to break the perfect bipartite 
symmetry of the system. This can be easily achieved with uniaxial strain \cite{Pereira2009}.

\begin{figure}[h]
\begin{center}
\includegraphics[width=1.0\columnwidth]{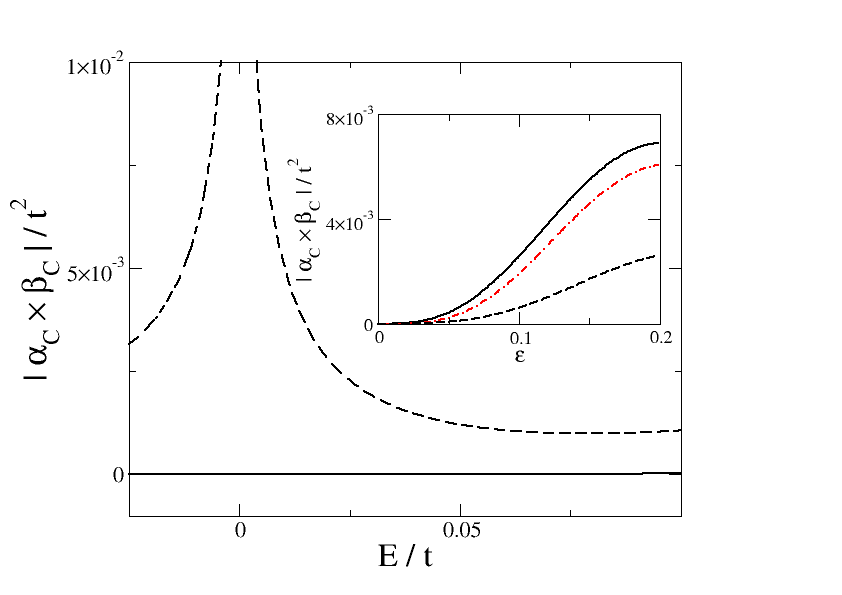}\\
\caption{$\vert \alpha_C \times \beta_C \vert$ as a function of Fermi energy for centre-bonded impurities. Solid (dashed) line corresponds 
to the case of unstrained (strained) impurities. Dashed line was obtained for a fixed value of strain, $\epsilon = 0.2$. 
Inset shows a log-scale plot of the difference between the two curves of the main panel plotted as a function of uniaxial strain for fixed values of Fermi energy. Solid line is for $E_F=0$, dot-dashed line is for $E_F=0.01 \, t$ and dashed line is for $E_F = 0.05 \,t$. }
\label{fig:interference2}
\end{center}
\end{figure}

To illustrate this point, we evaluate the quantity $\vert \alpha_C \times \beta_C \vert$ under 
the action of uniaxial strain and plot it as a function of the Fermi energy, shown by the dashed 
line in the main panel of Fig. \ref{fig:interference2}. For the sake of comparison, the solid line 
depicts the corresponding values for the strain-free case, as seen in Fig.~\ref{fig:interference}. 
The inset shows the difference between the strained and unstrained cases plotted as a function o
f the uniaxial strain $\epsilon$ for three different values of Fermi energy. Note that a small amount 
of strain is sufficient to destroy the interference seen in the cross section of graphene doped with 
centre-bonded impurities. 
This is a different manifestation of the underlying physics to the orbital symmetry discussion 
presented in Ref.~\onlinecite{Uchoa2011}.
Without the destructive interference in the scattering cross section the impurity transparency is 
lifted and centre-bonded dopants will act as strong scatterers just like the top-bonded counterparts. 
This is the ideal mechanism for sensitive strain sensors.  

In summary, we have shown that the bonding symmetry of impurities in graphene can tell whether 
they act as strong or weak scatterers, regardless of their specificity. In particular, impurities that 
are top-bonded to the underlying hexagonal lattice are the most suitable for being chemically 
sensed by graphene. In contrast, centre-bonded impurities in graphene are invisible to conduction 
electrons and unable to scatter them. Nevertheless, any mechanism that breaks the perfect 
hexagonal symmetry of centre-bonded impurities will lift this invisibility, causing a subsequent 
enhancement of the resistivity in these materials. Mechanical strain is one obvious mechanism, 
which suggests that graphene doped with centre-bonded impurities are ideal candidates for 
high-sensitivity strain sensors. Finally, despite the simplicity of our model, it is worth emphasizing
the generality of our finding. Having described the scattering of impurities through their self-energies, 
our conclusions are not dependent on specific choices of parameters but fundamentally dependent 
on symmetry arguments. We argue that by classifying dopants according to their bonding symmetry 
leads to a more efficient way of identifying strong and weak scatterers. Rather than trial and error, 
our approach offers a major advance to establish which substances is graphene a good sensor to. In the process of submitting our manuscript we learned of a work that obtains somewhat similar results by studying the conductivity corrections due to adsorbed impurities using the Kubo formula\cite{2016arXiv160600742R}. 

\acknowledgements
C.L. thanks the financial support of the Brazilian funding agencies CNPq and FAPERJ. 
The Ireland-based authors acknowledge financial support from the Programme for 
Research in Third Level Institutions (PRTLI). M.S.F. also acknowledges financial 
support from Science Foundation Ireland (Grant No. SFI 11/RFP.1/MTR/3083).

 \bibliography{RKKY-transport_resub}

\end{document}